\documentclass[a4paper,11pt]{article}
\usepackage{pos}

\usepackage{graphicx}
\usepackage{epstopdf}
\usepackage{subfigure}

\title{Computation of the Kugo-Ojima function from lattice simulations}
\ShortTitle{Kugo-Ojima function from lattice simulations}

\author[a]{Nuno Brito}
\author[a]{Orlando Oliveira}
\author*[a]{Paulo J. Silva}
\author[b]{Joannis Papavassiliou}
\author[b]{Mauricio N. Ferreira}
\author[c]{Arlene C. Aguilar}

\affiliation[a]{CFisUC, Department of Physics, University of Coimbra,  3004-516 Coimbra, Portugal}
\affiliation[b]{\mbox{Department of Theoretical Physics and IFIC, 
University of Valencia and CSIC}, \\
E-46100, Valencia, Spain}
\affiliation[c]{\mbox{University of Campinas - UNICAMP, Institute of Physics Gleb Wataghin,} \\
13083-859 Campinas, S\~{a}o Paulo, Brazil}

\emailAdd{nmrbrito2000@gmail.com}
\emailAdd{orlando@uc.pt}
\emailAdd{psilva@uc.pt}
\emailAdd{joannis.papavassiliou@uv.es}
\emailAdd{ansonar@uv.es}
\emailAdd{aguilar@ifi.unicamp.br}

\abstract{
In addition to its connection with a 
standard confinement criterion, the 
Kugo-Ojima function constitutes an indispensable component 
in a multitude of applications in the gauge sector of QCD.
In the present work we 
report on preliminary results 
of an ongoing  
large-volume lattice simulation of this 
special function. In particular, the volume-dependence 
of the data is studied in detail, 
and a comparison with 
results obtained from  
Schwinger-Dyson equations is carried out. 
}

\FullConference{The 40th International Symposium on Lattice Field Theory (Lattice 2023)\\
July 31st - August 4th, 2023\\
Fermi National Accelerator Laboratory\\}

\begin{document}
\maketitle

\section{Introduction}

The Kugo-Ojima (KO) function, 
$u(p^2)$, 
appears naturally in the context of 
the quantization  
formalism developed in~\cite{Kugo:1979gm,Kugo:1995km},  
where a non-trivial 
connection between confinement 
and the infrared behaviour of the gluon and ghost propagators 
in the Landau gauge was put forth. 
Within the KO formalism, the requirement of having
a well-defined BRST charge leads  
to the confinement criterion associated with the 
infrared behavior of $u(p^2)$, namely that $u(0) =-1$. It turns out that,  
in the Landau gauge, the realization of this condition would cause the divergence 
of the ghost dressing function, $F(p^2)$, at the origin~\cite{Kugo:1995km}, 
by virtue of the relation 
$F^{-1}(0) = 1+ u(0)$; for a review, see~\cite{Alkofer:2000wg}. However,  
as was established in a large number of 
works, the KO confinement condition is 
not fulfilled on the lattice
\cite{Nakajima:1999dq,Sternbeck:2006rd},   
and generally, in the context of the 
so-called "decoupling solutions", see, e.g.,~\cite{Ilgenfritz:2006he,Bogolubsky:2007ud,Cucchieri:2007md,Oliveira:2012eh} and~\cite{Aguilar:2008xm,Fischer:2008uz};
in particular, $u(0) \neq-1$, and 
$F(0) = c$, where $c$ is a finite constant.

The interest in the KO function 
resurged within the confines of the 
PT-BFM framework, namely the formalism that emerges 
from the fusion of the pinch technique (PT)~\cite{Binosi:2009qm} 
with the 
background 
field method (BFM)~\cite{Abbott:1980hw}. The 
relevance of $u(p^2)$ in this context  
originates from its coincidence with a 
central auxiliary function, denoted by 
$G(p^2)$ in the related literature
~\cite{Aguilar:2009pp}, i.e.\,,  
\begin{equation}
G(p^2) = u(p^2) \,.
\label{Gisu}
\end{equation}
The function $G(p^2)$ constitutes 
one of the cornerstones 
of the aforementioned framework, and 
is a central element 
in a multitude of theoretical relations 
and physical applications derived from it~\cite{Binosi:2002ez}. 
In particular,  
$G(p^2)$ is a common component of all 
formal identities relating the BFM correlation 
functions with those in the 
linear covariant gauges. The prime example 
of such an identity is the relation 
connecting the background and 
ordinary gluon propagators, 
$\widehat\Delta(p^2)$ and 
$\Delta(p^2)$, respectively, 
namely~\cite{Binosi:2002ez}
\begin{equation}
\Delta(p^2) = \widehat\Delta(p^2)[1+G(p^2)] \,,
\label{BQI}
\end{equation}
Thus, $G(p^2)$ 
constitutes a crucial ingredient in the 
determination of the effective interaction, 
${\cal I} (p^2)$, 
\begin{equation}
{\cal I}(p^2) = \alpha_s p^2 \Delta(p^2)[1+G(p^2)]^{-2} \,,
\label{eff}
\end{equation}
which is employed in the computation of hadronic observables~\cite{Binosi:2014aea}, where 
$\alpha_s = g^2/4\pi$.

In view of these considerations, in this work we present preliminary results of a new large-volume lattice  simulation 
of the KO function that is currently underway.

\section{Lattice procedure and setup}

The KO function $u(p^2)$
is defined as the scalar co-factor 
of the following two-point function 
of composite operators,
\begin{equation}
    \int d^4 x e^{i p(x-y)}\ \langle 0 | T \left( [(D_\mu^{a e} c^e(x)] [f^{b c d} A_\nu^d(y) \bar{c}^c(y)]\right)|0\rangle= \delta^{a b}\left(\delta^{\mu \nu}-\frac{p_\mu p_\nu}{p^2}\right) u(p^2) \,,
    \label{kobig}
   \end{equation}
where $D_\mu^{a e}$ 
is the covariant 
derivative in the adjoint 
representation, and $T$ 
denotes the standard time-ordering operation.

An explicit lattice definition for $u^{ab}(p^2) = \delta^{ab} u(p^2)$ can be given as follows
   \begin{equation}
     \mathcal{U}^{ab}_{\mu\nu}(p)=\frac{1}{V}\left\langle\sum_{x, y} \sum_{c, d, e} e^{-i p \cdot (x-y)} D_\mu^{a e}\left(M^{-1}\right)_{x y}^{e c} f^{b c d} A_{\nu}^d(y) \right\rangle_U \,,
     \label{kugolat}
   \end{equation}
where $M^{-1}$ is the ghost propagator and the scalar function $u(q^2)$ is given by
    \begin{equation}
        u(p^2) = \frac{1}{(N_d-1)(N_c^2-1)}\sum_{\mu,a}\mathcal{U}^{aa}_{\mu\mu}(p) \,.
    \end{equation}


In order to study the KO function on the lattice, we rely on Eq.~(\ref{kugolat}). However, for practical reasons, it is convenient to compute $\mathcal{U}^{ab}_{\mu\nu}(p)$ using a point source $y_0$ in the inversion
of the lattice Faddeev-Popov operator
   \begin{equation}
      \mathcal{U}^{ab}_{\mu\nu}(p)=\left\langle\sum_{x} \sum_{c, d, e} e^{-i p \cdot (x-y_0)} D_\mu^{a e}\left(M^{-1}\right)_{x y_0}^{e c} f^{b c d} A_{\nu}^d(y_0) \right\rangle_U \,.
        \end{equation}
The computation of the KO function on the lattice is performed following the procedure:
\begin{enumerate}
\item prepare the source, using a suitable lattice definition for   $f_{abc}A^{c}_{\mu}$:
 \begin{displaymath}
            f_{abc}A^{c}_{\mu}(x) = -\frac{1}{2}\text{Tr}\left[\left\{\left(U_{x,-\mu}^\dagger + U_{x,\mu}\right) - \left(U_{x,-\mu}^\dagger + U_{x,\mu}\right)^\dagger\right\}[t^a,t^b]\right] \, ;
  \end{displaymath}

\item solve the linear system of equations to get the ghost propagator, taking care of zero modes\newline
\begin{center}  $MY=M\phi_{b,\nu}$ \, ; \,
  $M \psi_{b,\nu}  = Y$ \, ;
\end{center}

\item apply the covariant derivative, which can be written on the lattice as \cite{Sternbeck:2006rd}
        \begin{displaymath}
            \left(D_\mu[U]\right)_{x y}^{a b}=2 \text{ Re} \operatorname{Tr}\left[t^b t^a U_{x, \mu}\right] \delta_{x+\hat{\mu}, y}-2 \text{ Re} \operatorname{Tr}\left[t^a t^b U_{x, \mu}\right] \delta_{x, y} \, ;
        \end{displaymath}
  
      \item apply a Fast Fourier Transform (FFT) and include the correction due to the location of the point source.
\end{enumerate}

In this work we consider quenched lattice ensembles generated with the Wilson gauge action, with  $\beta=6.0$ ($a\sim0.1$fm) for 
the lattice volumes $32^4$, $48^4$, $64^4$, and $80^4$, whose physical volumes go from $~(3\textrm{ fm})^4$ to $~(8\textrm{ fm})^4$. 
The results we show use 100 configurations for each of the lattice ensembles, except for the largest volume, where the number of gauge configurations
is 50. For the smallest lattice volume, we consider an average over several point sources; for the other volumes, only one point source is considered. Computer simulations have been performed with the help of Chroma~\cite{Edwards} and 
PFFT ~\cite{Pippig} libraries. 
The interested reader may find further details on the work reported here in~\cite{Brito:2023pqm}.

\section{Results}

\begin{figure}[t]
  \begin{center}
       \includegraphics[width=0.72\textwidth,angle=0]{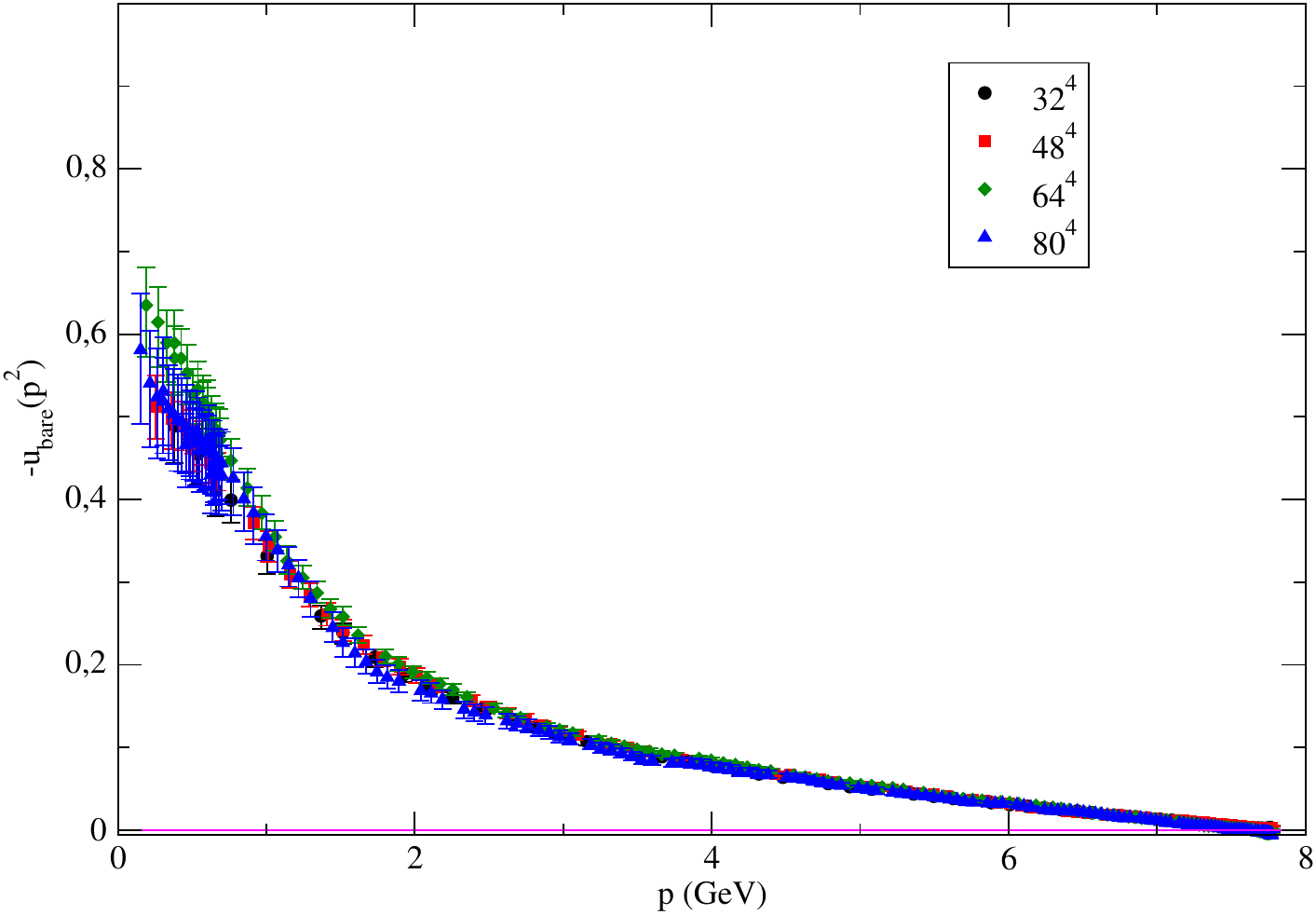}
  \end{center}
 \caption{Bare KO function for all lattice volumes. A horizontal line at $u_{bare}(p^2)=0$ is drawn to guide the eye.}
\label{kovol}
\end{figure}

The bare $u(p^2)$ for all lattice volumes is reported in Fig.~\ref{kovol}. 
The lattice data for the various volumes is compatible within errors, suggesting that the lattice volume effects are small or negligible. 

\begin{figure}[h] 
\vspace{0.55cm}
   \centering
   \subfigure[Longitudinal component.]{ \includegraphics[width=0.42\textwidth]{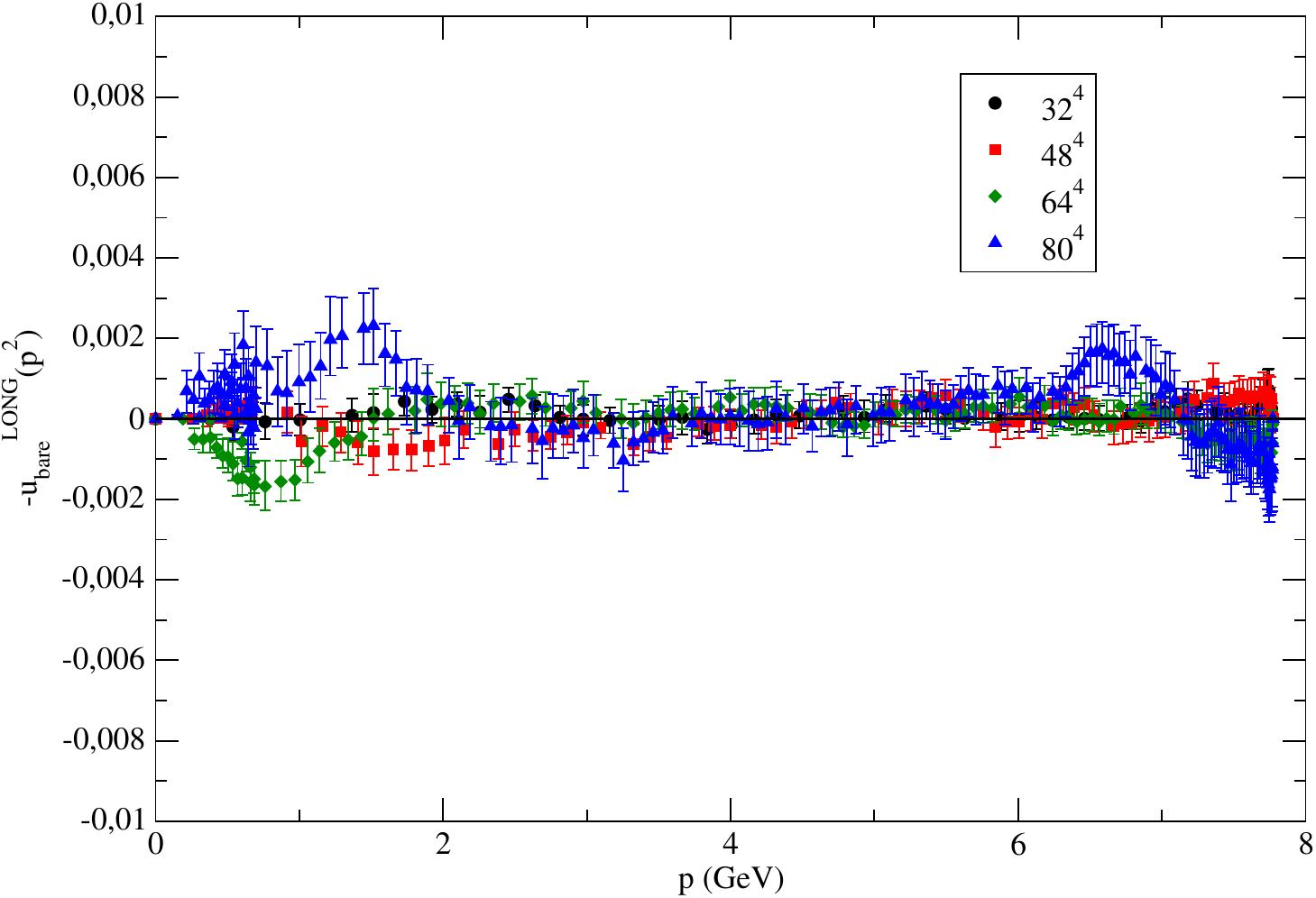} \label{kolongfig}} \qquad
   \subfigure[Imaginary part.]{ \includegraphics[width=0.42\textwidth]{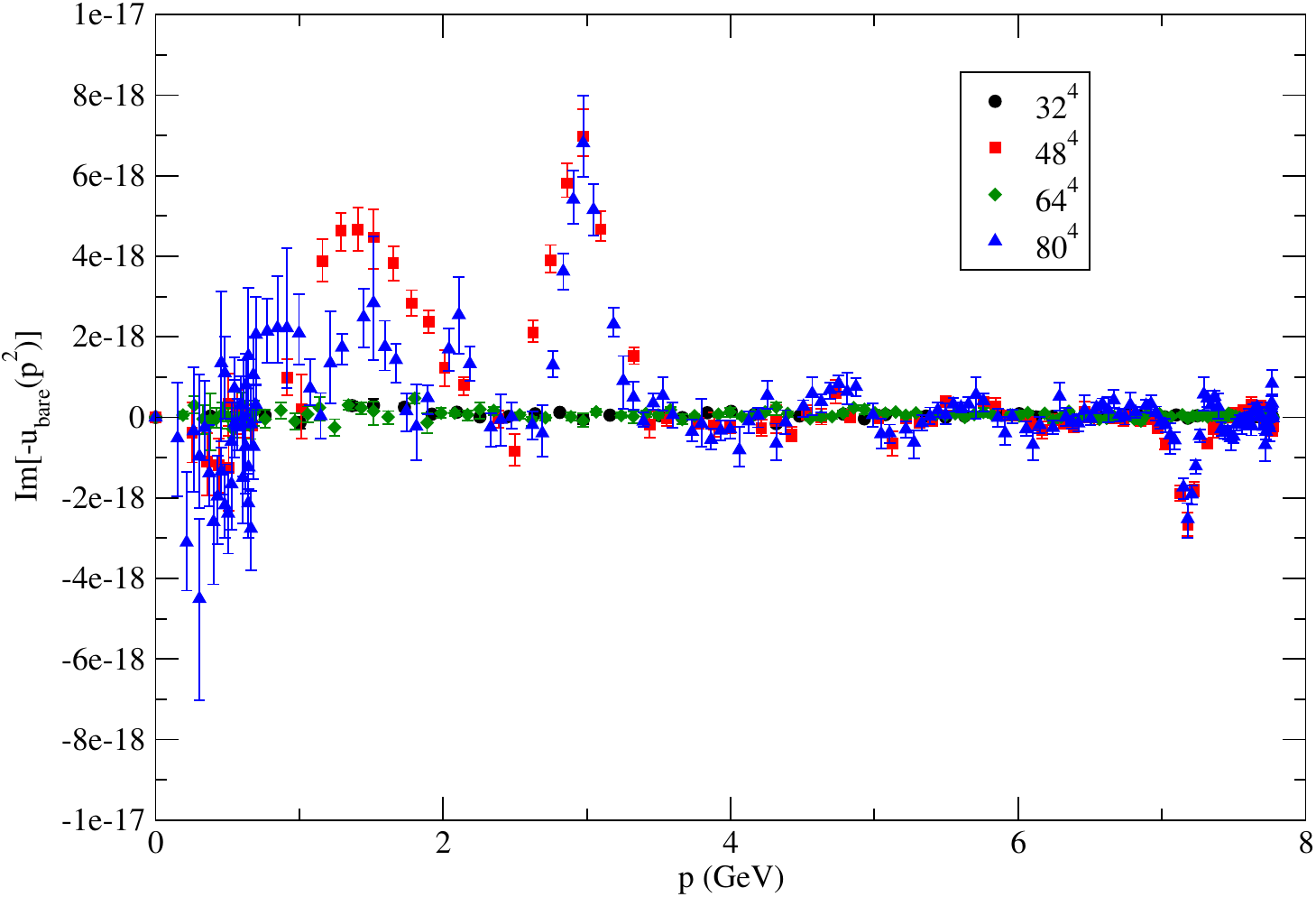} \label{koimagfig}}
  \caption{Longitudinal component and imaginary part of the Kugo-Ojima function.}
   \label{kolongimagfig}
\end{figure}

According to the definition in Eq.~(\ref{kobig}), the KO function is transverse. We have tested the transversality of the lattice KO function
computing its longitudinal projection that can be seen in Fig.~\ref{kolongfig}. Indeed, the numerical results show that the lattice version is, indeed,
orthogonal and that it can have a very small imaginary part that seems to decrease with an increase of the lattice volume.

\begin{figure}[h] 
   \centering
   \subfigure[Effect of number of sources, $32^4$ lattice.]{ \includegraphics[width=0.45\textwidth]{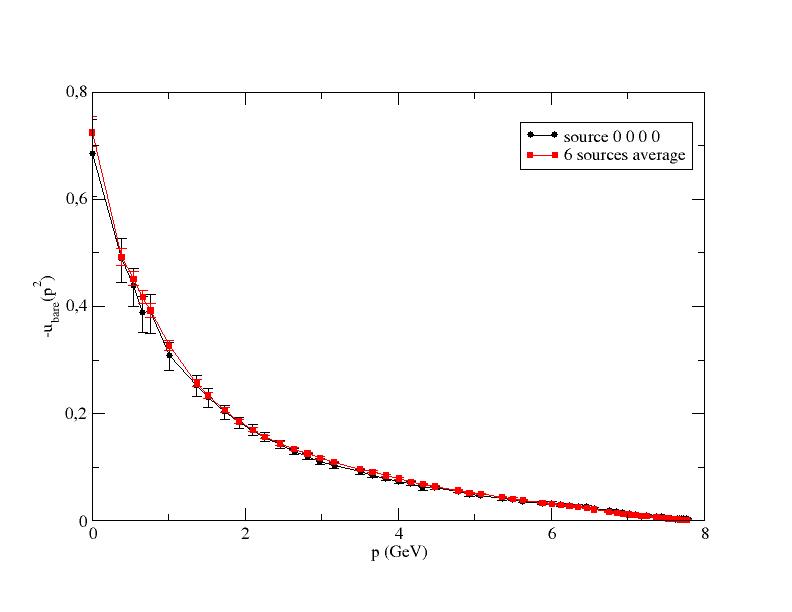} \label{sourcefig}} \qquad
   \subfigure[Momentum cuts, $64^4$ lattice.]{ \includegraphics[width=0.45\textwidth]{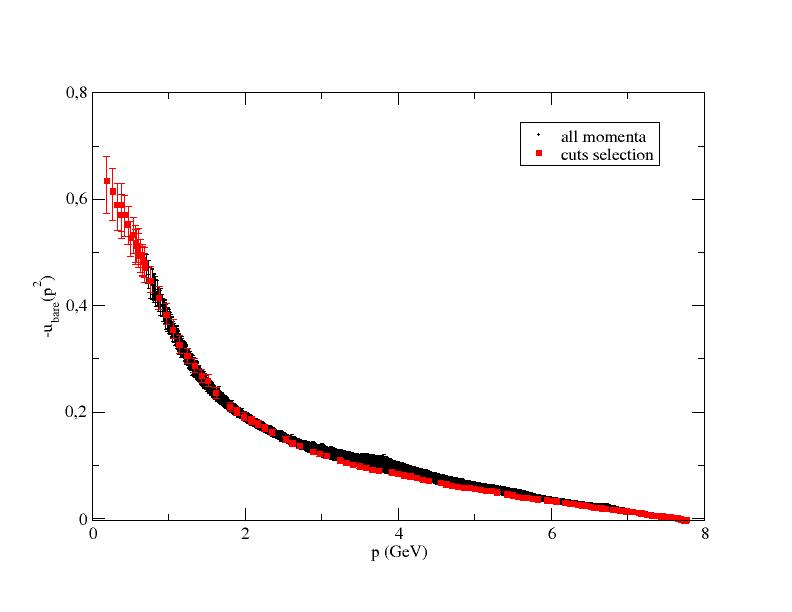} \label{cutsfig}}
  \caption{Other issues in the lattice computation of the KO function.}
   \label{statfig}
\end{figure}

For the smallest lattice volume, where we combined several point sources in the inversion of the Faddeev-Popov matrix, the effect of using
several sources is described in Fig.~\ref{sourcefig}. The curves compare the average over several point sources with the result computed with
a single point source at the origin of the lattice. The results show that averaging over several point sources reduces the fluctuations in $u(p^2)$.

\begin{figure}[h] 
   \centering
   \subfigure[$32^4$ lattice.]{ \includegraphics[width=0.45\textwidth]{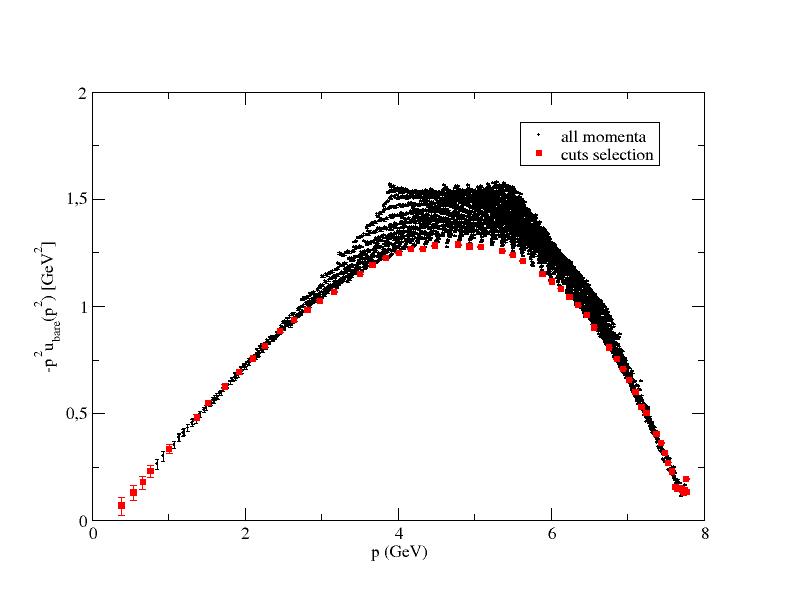} \label{onlydress}} \qquad
   \subfigure[$64^4$ lattice, H(4) extrapolation.]{ \includegraphics[width=0.45\textwidth]{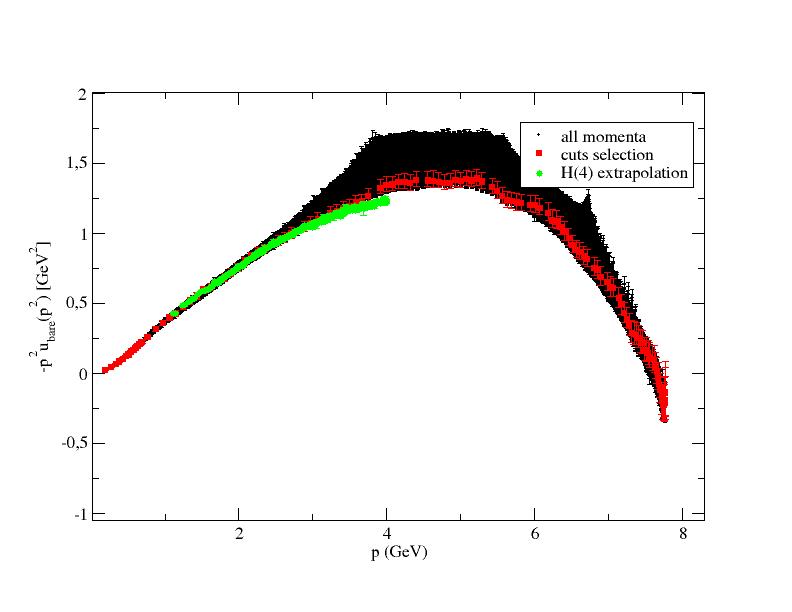} \label{h4ext}}
  \caption{Plots of $p^2 u(p^2)$.}
   \label{dress}
\end{figure}

The lattice data in the previous figures corresponds to the momenta that survives the cylindrical and conical cuts~\cite{Leinweber:1998uu}.
In Fig.~\ref{cutsfig} we compare the outcome of these cuts with all lattice data. 
Furthermore, in Fig.~\ref{onlydress} the product $p^2 u(p^2)$, which allows to better understand the breaking of rotational invariance on the lattice calculation of $u(p^2)$, is plotted. 
Fig.~\ref{h4ext} includes an H(4) extrapolation of the data that seems to provide good results for momenta below 3 GeV.

\begin{figure}[t]
  \begin{center}
    \includegraphics[width=0.72\textwidth,angle=0]{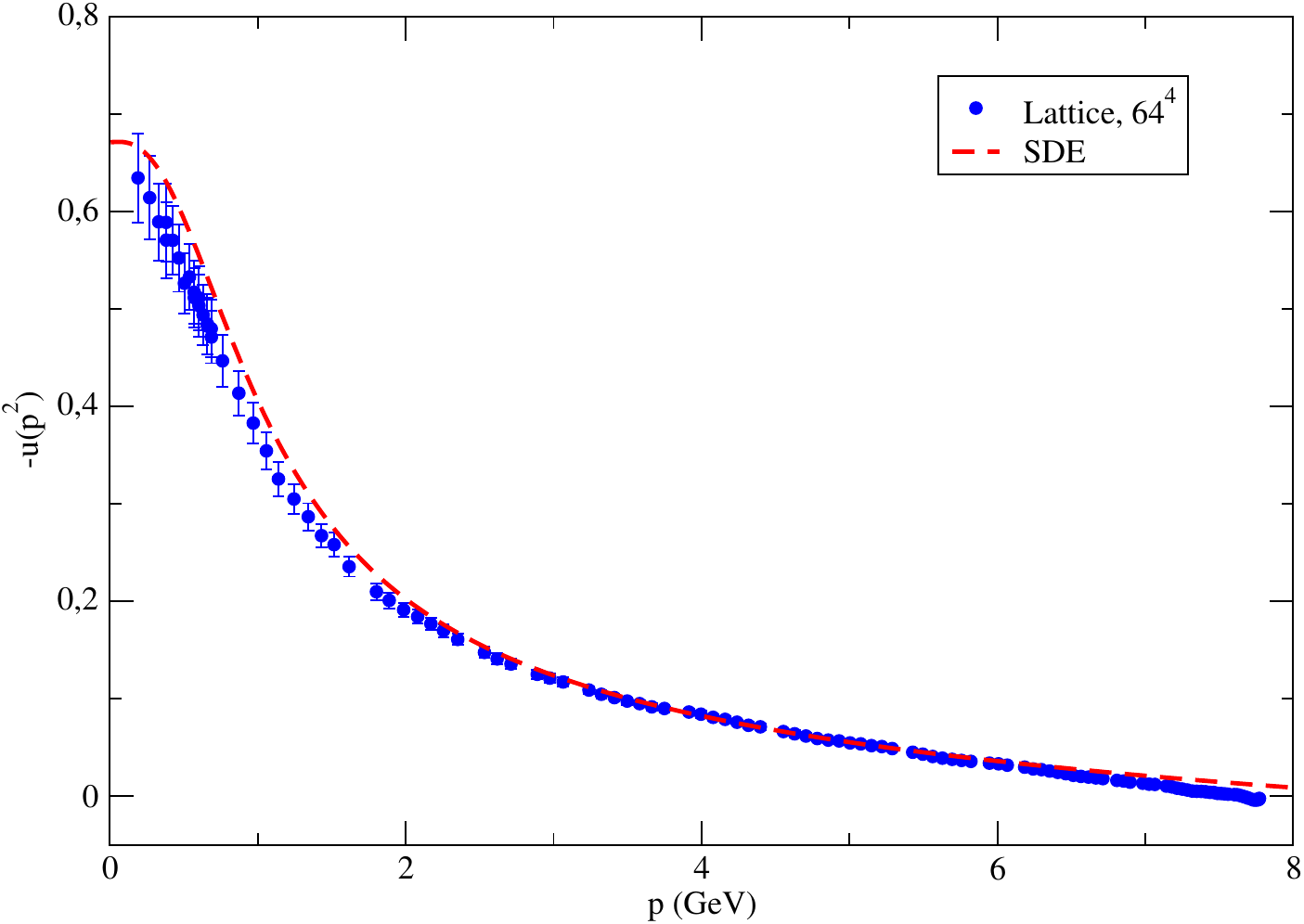}
  \end{center}
 \caption{Renormalized KO function and comparison with SDE results.}
\label{sde-ren}
\end{figure}

It is important to 
compare the results
of this simulation 
with those obtained from the 
Schwinger-Dyson equation (SDE) that 
governs the evolution of the function 
$G(p^2)$~\cite{Ferreira:2023fva}, and, therefore, 
by virtue of Eq.~(\ref{Gisu}), 
of $u(p^2)$. 
To that end, we renormalize  
the KO function for the simulation with the $64^4$ lattice by  
matching the lattice data with the outcome of an SDE calculation; the renormalization 
scale $\mu$ was chosen to be 
$\mu=4.3$ GeV. 
As can be seen in Fig.~\ref{sde-ren}, 
the renormalized lattice data and the 
SDE curve 
are in good agreement, especially 
in the region of momenta between 
$2-6$ GeV.

\section{Conclusions and Outlook}

In the present work we have reported recent 
results on the 
evaluation of the KO function on the lattice, for several lattice volumes. 
The results obtained are in agreement 
with those of earlier 
lattice studies~\cite{Sternbeck:2006rd} using smaller  volumes.
Moreover, they show good coincidence 
with the SDE results of~\cite{Ferreira:2023fva}.

We are currently increasing the statistics of our lattice ensembles, in order to reduce the errors in the deep infrared region; 
the final results will be 
reported elsewhere soon.
These results, 
in conjunction with existing lattice 
data for the gluon propagator, 
offer the possibility 
of deriving the effective interaction 
${\cal I}(p^2)$ of Eq.~(\ref{eff})
using exclusively ingredients from lattice QCD.

\section*{Acknowledgements}

\noindent

The authors acknowledge the computing time provided by the Laboratory for Advanced Computing at the University of Coimbra 
(FCT contracts  2021.09759.CPCA and  2022.15892.CPCA.A2). Work supported by FCT contracts  UIDB/04564/2020 (DOI \href{ https://doi.org/10.54499/UIDB/04564/2020}{10.54499/UIDB/04564/2020}), UIDP/\-04564/2020 (DOI \href{ https://doi.org/10.54499/UIDP/04564/2020}{10.54499/UIDP/04564/2020}) and CERN/FIS-PAR/0023/2021. P.~J.~S.  acknow\-led\-ges
financial support from FCT  contract CEECIND/00488/2017. A.~C.~A. is supported by the CNPq grant \mbox{307854/2019-1}, and acknowledges financial support from project 464898/2014-5 (INCT-FNA). M.~N.~F. and J.~P. are supported by the Spanish MICINN grant PID2020-113334GB-I00. M.~N.~F. acknowledges financial support from Generalitat Valenciana through contract \mbox{CIAPOS/2021/74}. J.~P. also acknowledges funding from the Generalitat Valenciana grant CIPROM/2022/66.


\end{document}